# Investigating the Emergence of Online Learning in Different Countries using the 5 W's and 1 H Approach

Nirmalya Thakur[1], Isabella Hall[1], Chia Y. Han[1]

[1] Department of Electrical Engineering and Computer Science, University of Cincinnati, Cincinnati, OH 45221-0030, U.S.A.

## ABSTRACT

The rise of the Internet of Everything lifestyle in the last decade has had a significant impact on the increased emergence and adoption of online learning in almost all countries across the world. E-learning 3.0 is expected to be-come the norm of learning globally in almost all sectors in the next few years. The pervasiveness of the Semantic Web powered by the Internet of Everything lifestyle is expected to play a huge role towards seamless and faster adoption of the emerging paradigms of E-learning 3.0. Therefore, this paper presents an exploratory study to analyze multimodal components of Semantic Web behavior data to investigate the emergence of online learning in different countries across the world. The work specifically involved investigating relevant web behavior data to interpret the 5 W's and 1 H - Who, What, When Where, Why, and How related to online learning. Based on studying the E-learning Index of 2021, the study was performed for all the countries that are member states of the Organization for Economic Cooperation and Development. The results presented and discussed help to interpret the emergence of online learning in each of these countries in terms of the associated public perceptions, queries, opinions, behaviors, and perspectives. Furthermore, to support research and development in this field, we have published the web behavior-based Big Data related to online learning that was mined for all these 38 countries, in the form of a dataset, which is available at https://dx.doi.org/10.21227/xbvs-0198.

# INTRODUCTION

The introduction of the paradigms of Web 1.0 (Fuchs et al. 2010), the first stage of the World Wide Web (W.W.W.), in and around 1989 (Choudhury 2014) led to the concept of making information and resources readily accessible on the internet. This was a significant progress that provided the platform for users across the globe to "connect" and see or read information seamlessly. This characteristic feature defined by the level of access led to the terminology "Read-Only-Web" getting associated with Web 1.0 since its inception (Richardson 2006). This technology was soon adopted by E-learning to expand its outreach. E-Learning 1.0 was characterized by its functionalities that enabled access to learning materials, opportunities, and resources to everyone connected on Web 1.0 (Jamali et al. 2017). These functionalities led to E-learning 1.0 receiving the tagline – "Anytime, Anywhere, and Anyone" (A.A.A.). The second generation of W.W.W., also referred to as Web 2.0 or the "Read-Write Web" in 2005 (w3.org 2020, Downes 2005), allowed users of W.W.W. to collaborate in addition to reading and viewing information. The E-learning industry was prompt to catch up with Web 2.0 by incorporating social aspects of learning theories in the E-learning platform, leading to the concept of E-learning 2.0 (Anderson 2020, Mödritscher 2006). Despite its features that helped people collaborate, there were a few limitations of Web 2.0, as mentioned in (Choudhury 2014). With an aim to address these limitations, the recent few years of W.W.W. have witnessed the development and emergence of Web 3.0 (lifeboat.com 2020). Web 3.0, also referred to as the "Executable Web," is centered around the concept of defining and linking data and its structures to facilitate effective and efficient discovery, automation, integration, interconnection, and reuse of data across a multitude of applications and platforms (Rudman et al. 2016). The E-learning industry has been one of the most rapidly advancing industries in the last decade and a half. The market for E-learning has increased by 900% since 2000 (Isaias 2019), and it is projected that the E-learning market will be worth $325 billion by 2025 (Chernev 2019). Recent research has also shown that E-learning increases retention rates by 25% to 60%. Society is, therefore, once again experiencing a paradigm shift in terms of E-learning, which is expected to lead to the wide-scale adoption of E-learning 3.0 across the globe. There have been multiple definitions of E-learning 3.0

that outline its multiple characteristics such as E-learning 3.0 = E-Learning 2.0 + Web 3.0 (Hussain 2012), E-learning = Anyone, Anywhere, Any-time, and A.I. (Rubens 2014), rhizomatic implications for blended learning (Cronje 2018), and several more (Aljawarneh 2020, Ivanova 2021). Therefore, it can be concluded that E-learning 3.0 is expected to be-come the norm in the Education Industry on a global scale in the next few years. Thus, investigating the emergence and rate of advancement of E-learning in different geographic regions of the world is of critical importance.

In the recent past, there have been several works in this field that have focused on multiple aspects of E-learning such as - investigating barriers to online learning (Baticulon et al. 2021), the efficacy of online learning courses (Castro et al. 2019), analyzing rates of student satisfaction in online learning environments (Landrum et al. 2021), studying perceptions of students towards online learning-based technologies (Cole et al. 2021), studying levels of student engagement in modern E-learning applications and platforms (Chiu et al. 2021), challenges faced by teachers and students related to these technologies (Hafeez 2021), and investigating the learning outcomes considering the effects of multiple factors (Yu 2021). However, none of these prior works have focused on investigating the emergence of online learning based on studying relevant web-behavior. Studying web behavior is fundamental towards understanding the emergence of E-learning 3.0 and the emerging advances in this field as advancements in W.W.W. and E-learning have historically been interrelated. Web behavior in today's Internet of Everything era (Chatzigiannakis et al. 2021) is characterized by the generation of tremendous amounts of Big Data. Due to its characteristics which include volume, variety, and veracity (Cappa et al. 2021), the mining and analysis of web behavior-based Big Data originating from different geographic regions of the world holds the potential for interpretation of the emergence of online learning in each of these regions. Addressing this research challenge by exploring the intersections of multiple disciplines serves as the main motivation for this research work.

Prior to performing such a Big Data-driven study, it is crucial to have a framework that helps to analyze the preparedness or readiness of a country to adopt E-learning and its emerging technologies in a seamless manner. Therefore, we referred to the framework discussed in (Preply 2021). This research (Preply 2021) focused on determining those attributes that could lead to a country being prepared for implementing digital and E-learning services and applications while also determining those set of implementation-level factors that contribute towards the success of online learning. The factors discussed in this work include (1) overall accessibility to online education, (2) amount of funds dedicated by a country for its education sector, (3) internet availability and speed, (4) cost of internet access, and (5) the E-learning climate that encompasses the topics of tutoring, market volume, and market growth. This paper is presented as follows. In Section 2, we present the steps that we followed for studying the emergence of Online Learning in all the 38 countries that are member states of the Organization for Economic Cooperation and Development (OECD). Section 3 presents the results and discussions, including a description of the dataset that we developed and published as a result of this research work. Section 4 concludes the paper by summarizing its main contributions and outlining the future scope of work in this field. It is followed by the author contributions section, which is followed by references.

## METHODOLOGY

To investigate the emergence of online learning in different countries of the world, we specifically focused on studying the emergence of online learning in all the 38 countries that are member states of the Organization for Economic Cooperation and Development (OECD) (OECD 2021). OECD is an intergovernmental organization dedicated to creating economic progress and increasing world trade. The organization consists of 38 member countries that all meet the criteria and description of having a high-income economy with a high human development index which then, in turn, has them considered as developed countries. The OECD provides in-depth knowledge and guidance to lesser developed countries concerning specific policies and socio-economic reforms (Kenton 2021, Wikipedia 2021). The OECD member countries collectively comprise 62.2 % of global nominal G.D.P. ($49.6 trillion) and 42.8 % of global G.D.P. ($54.2 trillion) at purchasing power parity (OECD 2021). Furthermore, the member states account for three-quarters of world trade, 63% of the world G.D., 95% of world official development assistance, more than 50% of the world's overall energy use, and roughly 18% of the total world population (Deitrs 1968). In view of these notable contributions of the member states of OECD, analyzing the emergence of online learning in these countries is expected to provide us an indication of the patterns of emergence of online learning in all the other countries of the world. Therefore, this study and the development of the dataset focused on investigating the Big Data-based web behavior related to online learning in all these countries.

There are multiple ways of mining, studying, and discovering patterns in Web Behavior-based Big Data (Salih et al. 2021). However, when it comes to studying E-learning based on web behavior that involves the user as a central and indispensable component, user-centered approaches for semantic and linguistic mining of Big Data can provide insight and support means of classification of the associated queries and searches for further investigation. The 5 W's model, primarily used in journalism, holds the potential to serve this purpose (Hart 1996). In journalism, the basic premise of writing re-volves around answering the five W's, which correspond to the basic questions that all audiences ask and can be categorized into one of the five categories. The five W's are 'who', 'what', 'when', 'where', and 'why'. Answering these five questions gives the audience a comprehensive collection of information regarding the event or situation to allow them to process what actually happened and determine whether or not they will be affected by it and, if so, to what extent. This process creates a "user-centered" information premise that corresponds to the user focusing on the extent of the impact on themselves and their lives. For the 'what', the things considered are what exactly is the problem or issue being addressed as well as what approach can be taken and what plan can be made to solve the given problem. For the 'who', the questions typically revolve around the user and who the user is responsible for in terms of determining the needs. The third W is 'where'; this focuses on helping the users on where to find information, especially in regard to performing specific tasks within a program or situation or event. The fourth W is 'when', which mainly focuses on telling the user the order of tasks that lead to a larger scenario in terms of time or con-text. The final W is 'why', and it explains the reasoning behind the implementation and need for each step in the order that they are in and need to be completed. User documentation is not complete without explaining the reasoning behind 'how' to solve specific problems or the implementations of the solutions.

Therefore, an extension of this model is popularly known as the 5W's and 1H model, which we applied to study the web behavior-based Big Data related to the emergence of online learning originating from different countries. We applied this 5 W and 1 H model to the Big Data-based web behavior related to E-learning originating from all the 38 member states of OECD. While such web behavior can be captured from various sources – analyzing web behavior from Google Search data holds the potential for maximum data availability as Google is the most popular search engine in the world, accounting for 70% of the market search share and capturing almost 85% of mobile traffic on a global scale (Forsey 2021). To perform the same, we used the intelligent online platform - answerthepublic.com (Answerthepublic 2021). Based on keyword comparison, this platform helps to capture the most frequent queries being asked in Google Search from different countries of the world on a specific day. The platform uses topic modeling, text mining, content parsing, and a host of other Natural Language Processing and Artificial Intelligence-based methodologies to perform the same. These queries could potentially include any forms of questions asked by users while working or browsing the internet. Answer The Public helps the user mine the relevant Big Data related to Google searches and the associated metadata in a faster and efficient manner as the platform helps to structure and classify the data into different categories to make the data easier to study and interpret. This structured data is usually in the form of multiple queries captured from Google Searches, which are classified as per the binning approach of Big Data mining (Alotaibi et al. 2016). For the purpose of this research, we focused on applying data filters to consider all the queries with a specific focus on the 5W's and 1H in the context of the E-learning-based Big Data that we mined for all these 38 countries on November 1, 2021. This date was chosen as it was the most recent date for which this data could be mined at the time of submission of this paper.

## RESULTS AND DISCUSSIONS

As mentioned in Section 2, we performed this study using the platform – Answer The Public by referring to the framework presented in (Hart 1996) and investigating the 5W's and 1H of all queries on Google that originated from all the 38 member states of OECD on November 1, 2021. For each of these countries, the set of keywords that were provided to this platform were – "online learning," and thereafter, the associated queries were captured and filtered out to retain only those queries that consisted of the modifiers associated with the 5W's and 1H. The data for all these countries were collected and analyzed on November 1, 2021. Even though we specifically focus on investigating the 5 W's and 1 H on E-Learning, to support research and development in this field, we mined all the questions and queries (including questions with 5 W and 1 H) for all these 38 countries and compiled the Big Data in the form of a dataset which is available at https://dx.doi.org/10.21227/xbvs-0198.

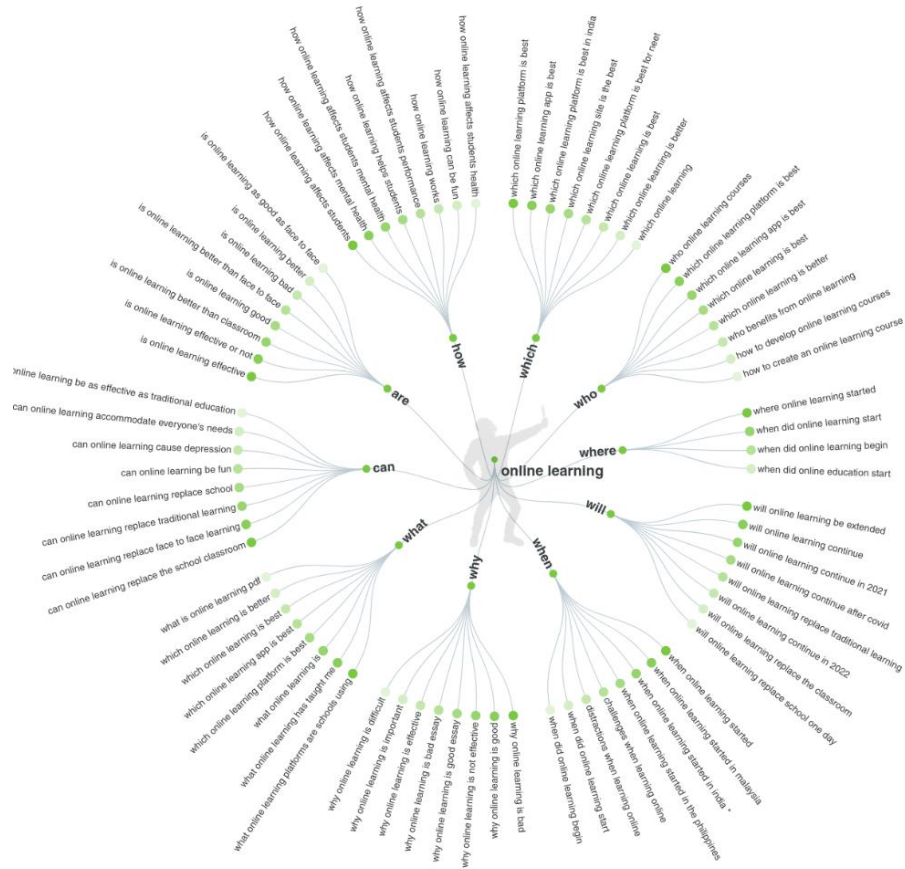

Figure 1. Result of mining and analyzing web behavior-based Big Data and categorizing the queries into specific labels, including the 5 W's and 1H for Australia

The dataset is presented in the form of one M.S. Excel Workbook that consists of the data of all these 38 countries in the form of different M.S. Excel Sheets. To mine, interpret, and analyze the data for each of these 38 countries, the steps mentioned in Section 3 were repeated in a sequential manner. This analysis of one of these countries, Australia, is shown in Figure 1. As can be seen from Figure 1, the search queries were classified into bins or categories (representing the 5 W's and 1H as well as other categories) to group together similar questions related to online learning based on the study of relevant Google search-based web behavior data. For paucity of space, we could not provide the results representing this classification for the remaining 37 countries in this paper. However, the data has been compiled for all these countries and is presented in this dataset to support further research and development in this field.

## CONCLUSIONS

The E-learning industry, one of the largest and fastest-growing industries on a global scale, has experienced several developments and the emergence of technologies in the last decade and a half. Now the society is advancing towards wide-scale acceptance of the most emerging form of E-learning, E-learning 3.0, which is expected to become the norm in almost all the geographic regions of the world in the next few years. The extensive use of the modern-day World Wide Web in today's Internet of Everything era, leading to the generation of multimodal forms of web behavior-based Big Data, holds the potential for analysis of the patterns of emergence of E-learning 3.0 in different countries of the world by mining and interpretation of the relevant web behavior-based data. Therefore, in this paper, we present a user-centered approach by applying the 5W-1H model, a popular model in journalism, to identify queries in terms of the 5 W's and 1 H - Who, What, When Where, Why, and How, and beyond, related to online learning that emerged from different countries of the world. The work involved exploring the intersections of Internet of Everything, Big Data, Natural Language Processing, Artificial Intelligence, and Human-Computer Interaction. As online learning and its emergence is a crucial factor for the economic growth and development of a nation, so our study focused on investigating the 5W's and 1H related to online learning for all the 38 countries that are member states of the Organization for Economic Cooperation and Development (OECD). The results presented outline the potential and relevance of this approach for the interpretation of the emergence of online learning in each of these countries through the lens of the 5W-1H model. Furthermore, to advance research and development in this field, we mined all forms of queries related to online learning (including the 5W's and 1H) and developed a dataset that may be used for the investigation of similar research questions in this field of research. Future work would involve expanding this dataset to include all the countries of the world to further support its applications and use cases in advancing research and development in this field.

# AUTHOR CONTRIBUTIONS

Conceptualization, N.T.; Methodology, N.T.; Data Curation, N.T. and I.H.; Formal Analysis, N.T. and I.H; Data Visualization and Interpretation, N.T and I.H.; Results, N.T; Writing-Original Draft Preparation, N.T. and I.H; Writing-Review and Editing, N.T; supervision, N.T; project administration, N.T and C.Y.H.; Funding Acquisition, Not Applicable. All authors have read and agreed to the published version of the manuscript.